\documentclass{nojournal}
\usepackage{url} 

\journalname{Journal of Geophysical Research: Oceans}

\begin{document}

%
%

\title{The Impact of a Reduced High-wind Charnock Parameter on Wave Growth With Application to the North Sea, the Norwegian Sea and the Arctic Ocean\footnote{Accepted for publication in \textit{J. Geophys Res: Oceans}, 2022-03-10, DOI:10.1029/2021JC018196}}

%
%



\authors{{\O}yvind Breivik\affil{1,4}, Ana Carrasco\affil{1}, Hilde Haakenstad\affil{1,4}, Ole Johan Aarnes\affil{1}, Arno Behrens\affil{3}, Jean-Raymond Bidlot\affil{2}, Jan-Victor Bj{\"{o}}rkqvist\affil{1,5}, Patrik Bohlinger\affil{1}, Birgitte R Furevik\affil{1,4}, Joanna Staneva\affil{3}, and Magnar Reistad\affil{1}}

\affiliation{1}{Norwegian Meteorological Institute, Norway}
\affiliation{2}{European Centre for Medium-Range Weather Forecasts, United Kingdom}
\affiliation{3}{Helmholtz-Zentrum Hereon, Germany}
\affiliation{4}{University of Bergen, Norway}
\affiliation{5}{Finnish Meteorological Institute, Finland}




\correspondingauthor{{\O}yvind Breivik}{oyvind.breivik@met.no, ORCID: \texttt{0000-0002-2900-8458}.}




\begin{keypoints}
\item A semi-empirical parameterization for high-wind drag (above 30 m~s$^{-1}$) is implemented in WAM
\item A two-year comparison against a control run shows significant bias reduction in strong winds
\item A 23-year wave hindcast shows good performance compared to satellite and buoy measurements

\end{keypoints}

%
%

\begin{abstract}
As atmospheric models move to higher resolution and resolve smaller scales, the maximum modeled wind speed also tends to increase. Wave models tuned to coarser wind fields tend to overestimate the wave growth under strong winds. A recently developed semi-empirical parameterization of the Charnock parameter, which controls the roughness length over surface waves, substantially reduces the aerodynamic drag of waves in high winds (above a threshold of 30 m~s$^{-1}$). Here we apply the formulation in a recent version of the wave model WAM (Cycle 4.7), which uses a modified version of the physics parameterizations by \citet{ardhuin10} as well as subgrid obstructions for better performance around complex topography. The new Charnock formulation is tested with wind forcing from NORA3, a recently completed non-hydrostatic atmospheric downscaling of the global reanalysis ERA5 for the North Sea, the Norwegian Sea and the Barents Sea. Such high-resolution atmospheric model integrations tend to have stronger (and more realistic) upper-percentile winds than what is typically found in coarser atmospheric models. A two-year comparison  (2011--2012)  of a control run against the run with the modified Charnock parameter shows a dramatic reduction of the wave height bias in high-wind cases. The added computational cost of the new physics and the reduction of the Charnock parameter compared to the earlier WAM physics is modest (14\%). A longer (1998--2020) hindcast integration with the new Charnock parameter is found to compare well against in situ and altimeter wave measurements both for intermediate and high sea states.
\end{abstract}

\section*{Plain Language Summary}
Wave models are sensitive to strong winds, and as the atmospheric models have increased in resolution, the strength of the winds has also increased as small-scale features of synoptic storms become more realistically modeled. 
Here we investigate the behavior of a bespoke version of the wave model WAM where we have modified how waves grow in strong winds. More specifically, we have modified the so-called Charnock parameter, which determines how rough the sea surface gets as the wind picks up. By making the sea surface smoother under strong winds we reduce the growth of the waves under hurricane conditions in a manner which appears to be in line with recent studies of the behavior of the sea surface in very strong winds. The results match our observations very well over a wide range of conditions throughout the model domain and show a clear improvement over a control experiment with the same wave model without a modified Charnock parameter. Finally, a detailed wave hindcast, or sea state archive, for the Norwegian Sea, North Sea and the Arctic Ocean covering the period 1998--2020 is presented where this modified Charnock parameter is employed. The results compare well against observations. 

\section{Introduction}
\label{sec:intro}
Third generation spectral wave models \citep{has88,tol91,kom94,boo99,jan04,tolman14,wam47r1} balance the energy and momentum flux from the wind field against the dissipation due to wave breaking \citep{jan04} as well as the nonlinear transfer of energy within the wave spectrum. Modern wave models tend to perform well in terms of integrated wave parameters such as significant wave height, mean wave period and mean wave direction as long as the quality of the wind field is high (see, e.g., \citealt{hersbachetal2020}). It is however not clear whether wave models also perform well under very strong winds \citep{jensen06,du17wblm}, and it is known that the wind input source term based on conservation of momentum in the boundary layer first proposed by \citet{janssen89} tends to overestimate the drag at high winds \citep{jensen06,du17wblm}.

The growth of waves under extratropical wind storms and tropical cyclones has been the topic of several studies in the past two decades \citep{powell03,donelan04,zweers10,chen13directional}. Wave growth is controlled by the aerodynamic roughness of the surface, i.e., the drag that is felt by the wind. There is increasing evidence from theoretical \citep{makin05}, laboratory \citep{donelan04, curcic2020} and field studies \citep{powell03,holthuijsen12hurricanewaves,donelan18} that the roughness (and thus the drag) starts to level off or even drop \citep{powell03} at very high wind speeds. The exact threshold where the drag will start to level off remains unclear, with \citet{curcic2020}, in a recent revisit of the drag saturation rate measured by \citet{donelan04}, suggesting that it may start as low as around 20~m~s$^{-1}$ (see also \citealt{bi2015}). Other studies (see the overview in Fig. 6 by \citealt{holthuijsen12hurricanewaves}) report a levelling off from around 30~m~s$^{-1}$ \citep{zweers10}. 

As the resolution of atmospheric models continues to increase and the models increasingly include non-hydrostatic physics, the intensity of the strongest wind storms modeled also increases. This does not mean that the winds are overestimated, merely that high-resolution non-hydrostatic atmosphere models capture small-scale features that cannot be resolved by coarser models \citep{haakenstad21nora3}. 
Since wave models have hitherto not realistically modeled the reduction in drag under the strongest winds \citep{du17wblm}, this tendency for coarser, hydrostatic atmospheric models to produce too weak winds may have been partly compensated by the wind input source term in the wave model. 
Thus, how wave models parameterize wave growth under high winds becomes increasingly important as forecast systems, both coupled and uncoupled, move towards higher resolution. That is the topic of this paper.

There are several semi-empirical approaches to reducing the momentum flux to the wave field under  strong winds.  A parameterization for the direct reduction of the drag coefficient was proposed by \citet{holthuijsen12hurricanewaves}. As summarised by \citet{du17wblm}, other approaches include capping the limit on the ratio between the wind speed and the friction velocity \citep{jensen06} and setting a limit on the roughness length \citep{ardhuin10}. A spectral sheltering to reduce the high-frequency wave growth in the presence of longer waves was formulated by \citet{banner10}, following work by \citet{chen00}. The impact of limiting the maximum steepness of short waves was reported by \citet{magnusson19ecmwf}. In addition, the Charnock parameter itself \citep{charnock55,janssen89} can be modified for strong winds, thus directly controlling the wave growth by limiting the surface roughness. This is an approach somewhat similar to the cap on the roughness length introduced by \citet{ardhuin10}. This modification of the Charnock parameter is specifically investigated in this paper. 

This article is laid out as follows. We first present in Section~\ref{sec:wam47} the relevant details of the wind input source term as implemented in the WAM Cycle 4.7 wave model, a recent version of the open-source WAM model \citep{has88,gun92,kom94}. In Section~\ref{sec:hidrag}  we present a recently proposed semi-empirical reduction of the Charnock parameter at high winds \citep{wam47r1,li2021improving}.
Section~\ref{sec:wamruns} presents a comparison of two wave model runs, one with a reduced Charnock parameter at high winds and a control run with no reduction, covering a two-year period. 
Section~\ref{sec:hindcast} presents a 23-year (uncoupled) wave hindcast where we have applied the new Charnock parameter to a WAM domain forced with high-resolution winds (3~km) from the recent NORA3 atmospheric hindcast \citep{haakenstad21nora3}, hereafter referred to as the NORA3 wave hindcast or simply NORA3 WAM. Finally, in Section~\ref{sec:discussion}, we summarise our findings and draw some conclusions regarding the usefulness of implementing a reduced-drag parameterization for high-resolution forecast and hindcast systems, both coupled and uncoupled.

\section{The wind input source term in WAM Cycle 4.7}
\label{sec:wam47}
The WAM wave model \citep{has88,gun92,kom94} has in recent years undergone major code restructuring as it has been made openly accessible through the EU project MyWave \citep{behrens13}. WAM Cycle 4.7 contains model physics that is similar to what is described by \citet{ardhuin10} and often referred to as ``Source term package 4'' (ST4), but with some differences indicated below. The implementation of the source terms is taken from a recent version of the wave model component (ECWAM Cycle  47R1, see \citealt{wam47r1}) of the Integrated Forecast System (IFS) operated by the European Centre for Medium-Range Weather Forecasts (ECMWF).

The air-side stress, $\tau_\mathrm{a} \, [\mathrm{Pa}]$, is supported almost entirely by the roughness of the oceanic wave field itself \citep{janssen89,janssen91}. There must thus exist a relationship between the wind stress and the roughness length $z_0$ of a water surface with waves, as \citet{charnock55} showed,
\begin{equation}
    z_0 = \alpha u_*^2/g.
    \label{eq:charnock}
\end{equation}
Here, $\alpha$ is a dimensionless coefficient known as the Charnock parameter and $g = 9.81\,\mathrm{m\,s^{-2}}$ is the gravitational acceleration. The   
friction velocity $u_*$ relates to the air-side stress $\tau_\mathrm{a}$ as 
\begin{equation}
    u_* \equiv \sqrt{\tau_\mathrm{a}/\rho_\mathrm{a}}
    \label{eq:ustar}
\end{equation}
with $\rho_\mathrm{a}$ the atmospheric density, assumed here to be constantly $1.225 \,\mathrm{kg\,m^{-3}}$. 
The dimensionless drag coefficient $C_\mathrm{d}$ is a bulk parameter as it relates an atmospheric state variable (the wind speed) at a specific height ($z = 10\,\mathrm{m}$) to a momentum flux,
\begin{equation}
    \tau_\mathrm{a} = \rho_\mathrm{a}u_*^2 = \rho_\mathrm{a}C_\mathrm{d}U_{10}^2.
    \label{eq:drag}
\end{equation}
The drag coefficient has been the subject of many studies over the years, and we refer the reader to \citet{edson13} for a thorough overview of the drag coefficient parameterizations currently in use. 
The drag is related to the roughness length, Eq.~(\ref{eq:charnock}), through the logarithmic wind profile,  which itself can be derived from a dimensional argument that in a constant-flux layer near the surface [see, e.g., \citealt{stu88}]  there is no divergence of momentum and hence no change in wind speed with time. The wind profile must thus be logarithmic,
\begin{equation}
    U(z) = \frac{u_*}{\kappa}\ln((z+z_0)/z_0), \, z > z_0.
    \label{eq:logprofile}
\end{equation}
Here, $\kappa \approx 0.4$ is von K{\'a}rm{\'a}n's constant. 
This eventually leads to the relation between the drag coefficient and the roughness length,
\begin{equation}
    C_\mathrm{d} = \frac{\kappa^2}{\ln^2((z+z_0)/z_0)}.
    \label{eq:drag_vs_roughness}
\end{equation}

We will now start with the observation, made by \citet{janssen89,janssen91}, that the Charnock parameter $\alpha$ in Eq.~(\ref{eq:charnock}) is not a constant, but is in fact a function of the sea state,
\begin{equation}
    \alpha = \frac{\hat{\alpha}}{\sqrt{1-\tau_\mathrm{in}/\tau_\mathrm{a}}}.
    \label{eq:varcharnock}
\end{equation}
Here $\tau_\mathrm{in}$ is the momentum flux to the wave field and is directly related to wave growth and $\hat{\alpha}$ is a parameter which, as was noted already by \citet{janssen89}, is controlled by the shortest gravity-capillary waves. It is commonly taken to be a constant and henceforth referred to as the minimum Charnock parameter since it represents the lowest value that $\alpha$ can attain. Since Cycle 47R1 of the ECWMF WAM model (ECWAM, see \citealt{wam47r1}) it has been kept at $\hat{\alpha} = \hat{\alpha}_0 = 0.0065$. 

It is clear that when waves absorb a sizeable amount of the momentum (when they grow quickly), the denominator in Eq.~(\ref{eq:varcharnock}) becomes small, making the Charnock parameter large. This means that in strong winds, where waves are young and growing, the drag will become large. 
This has dramatic consequences for the wave growth in standalone wave models with no feedback to the atmospheric model, but also for coupled systems, such as the IFS, where too much drag lowers the near-surface wind speed excessively in storm conditions \citep{pineauguillou20,bidlot20,li2021improving}.

The wave growth is controlled by the wind input term $S_\mathrm{in}$. The form used here is based on the formulation presented as Eq.~(19) by \citet{ardhuin10}. We repeat it here for convenience,
\begin{equation}
    S_\mathrm{in} = \frac{\rho_\mathrm{a}\beta_\mathrm{max}}{\rho_\mathrm{w}\kappa^2} \mathrm{e}^Z Z^4\left[\frac{u_*}{c}\right]^2 \mbox{max}\left(\cos(\theta-\phi), 0\right)^p\sigma F(k,\theta). 
    \label{eq:st4}
\end{equation}
Here, $F(k,\theta)\, [\mathrm{m^3\,rad^{-1}}]$ is the wave variance density in wavenumber ($k$)-direction ($\theta$) space, $\phi$ is the wind direction, $\beta_\mathrm{max}$ is a constant nondimensional growth parameter and
\begin{equation}
    Z = \ln{(kz_1)} + \kappa/\left[\cos(\theta-\phi)(u_*/c+z_\alpha)\right]
\end{equation}
is an effective wave age with $c$  the phase speed, $\sigma$  the intrinsic circular frequency $\mathrm{[rad\,s^{-1}]}$ and $z_\alpha$  a dimensionless wave age tuning parameter that shifts the growth curve. The directional spread is controlled by the power $p$, a tunable constant which is commonly (and  here) set to 2. Higher powers give a more narrowly directed wind input.

It is important to note that \citet{ardhuin10} already introduced a cap on the surface roughness in the form 
\begin{equation}
    z_0 = \min(\alpha_0u_*^2/g, z_\mathrm{0,max})
\end{equation}
which in turn is used to calculate 
\begin{equation}
    z_1 = \frac{z_0}{\sqrt{1-\tau_\mathrm{in}/\tau}}.
\end{equation}
\citet{ardhuin10} mention that imposing $z_\mathrm{0,max} = 0.0015$ corresponds to capping the drag coefficient at $C_\mathrm{D} = 2.5\times 10^{-3}$. It is thus clear that adjusting $\beta_\mathrm{max}$ and $z_\mathrm{0,max}$ allows some freedom in tuning a wave model to an atmospheric model, but they will not allow a reduction of the roughness length above a certain  wind speed threshold. It is also important to note that $z_\mathrm{0,max}$ is not always active (i.e., it is set to 1). This is the case for the TEST471 tuning, considered the best option for global wind fields (see Table 2.6 by \citealt{tolman19}). We have not used the $z_\mathrm{0,max}$ parameter in our WAM implementation.

A further tuning parameter introduced by \citet{ardhuin10} is a wavenumber-dependent sheltering effect where $u_*^2$ in Eq.~(\ref{eq:st4}) is replaced by 
\begin{equation}
   ({u'_*}^2)_i = u_*^2(\cos \phi, \sin \phi) - \tau_\mathrm{shelter} \frac{\rho_\mathrm{w}}{\rho_\mathrm{a}}g\int_0^{2\pi}\!\int_0^k \frac{k'}{\omega} S_\mathrm{in} (\cos \theta, \sin \theta) \, \mathrm{d}k'\, \mathrm{d}\theta.
   \label{eq:shelter}
\end{equation}
Here, $\omega = 2\pi f\,\mathrm{[rad\,s^{-1}]}$ is the circular frequency (identical to the intrinsic frequency $\sigma$ in the absence of currents) and $\rho_\mathrm{w}$ is the density of sea water, assumed here to be constantly $1000\,\mathrm{kg\,m^{-3}}$. Note also that the equation is applied in vector form with $\phi$ the wind direction and subscript $i$ indicating $x$ and $y$ components.
This sheltering effect also yields somewhat weaker growth in high-wind situations. The sheltering coefficient can vary between $0$ and $1$. Here we use a rather modest sheltering with $\tau_\mathrm{shelter} = 0.25$.

ECWAM Cycle 47R1 and WAM Cycle 4.7 also impose a maximum steepness for the high-frequency part of the spectrum \citep{magnusson19ecmwf,wam47r1} by demanding the spectrum be limited to
\begin{equation}
    F(f,\theta) = \min(F, F_\mathrm{max}).
\end{equation}
Here, 
\begin{equation}
    F_\mathrm{max} = \frac{\alpha_\mathrm{max}}{\pi}g^2(2\pi)^{-4}f^{-5}
    \label{eq:phillips}
\end{equation}
is the Phillips spectrum \citep{phillips58} with an omni-directional normalizing constant $(2\pi)^{-4}$ and the linear frequency $f=\omega/2\pi$. This maximum steepness is a plausible limiting mechanism for the roughness length in high-wind situations as short waves should start to break as they get steeper than what the Phillips spectrum dictates. We have set $\alpha_\mathrm{max} = 0.031$ in accordance with ECWAM \citep{wam47r1}.

\section{A semi-empirical reduction of the Charnock parameter in high-wind regimes}
\label{sec:hidrag}
The drag coefficient (for wind at $z=10\,\mathrm{m}$) can be seen from Eq.~(\ref{eq:drag_vs_roughness}) to be related to the roughness length of the sea surface as
\begin{equation}
    z_0 = (10+z_0)\exp{(-\kappa/\sqrt{C_\mathrm{d}})} \approx 10\exp{(-\kappa/\sqrt{C_\mathrm{d}})}.
    \label{eq:z0}
\end{equation}
Since the wave growth depends on the momentum flux it is closely linked to the drag coefficient, see Eq.~(\ref{eq:drag}). For varying wave ages $c/u_*$, the ratio $\tau_\mathrm{in}/\tau_\mathrm{a}$ in Eq.~(\ref{eq:varcharnock}) will also vary. The roughness is controlled by both long and short waves. The roughness of long waves is controlled by the denominator in Eq.~(\ref{eq:varcharnock}). Young sea will tend to have high ratios since the wave growth is rapid. This in turn leads to a small denominator in Eq.~(\ref{eq:varcharnock}) and high Charnock values. This is the impact of the long (resolved) waves on the surface roughness. As mentioned before it is clear that the wave growth will be stronger for young wind sea than for older wind sea, and for really strong winds, approaching 30~m~s$^{-1}$, the sea state is always young as such strong winds are rarely sustained over long periods \citep{li2021improving}. If the roughness does indeed go down for very high winds, it seems reasonable to attempt to adjust $\hat{\alpha}$ since, as \citet{janssen89} pointed out, this parameter represents the roughness due to the shortest (unresolved) waves of the $f^{-5}$ Phillips tail, Eq.~(\ref{eq:phillips}). In essence this will allow us to control the roughness of the shortest waves for different wind regimes.

\citet{li2021improving}, employing the recent implementation from ECWAM \citep{wam47r1}, tested the following minimum Charnock parameterization for typhoon Lingling in a coupled atmosphere-wave model run, 
\begin{equation}
    \hat{\alpha} = \hat{\alpha}_{-} + 0.5\left(\hat{\alpha}_0- \hat{\alpha}_{-}) (1-\tanh{[(U_{10}-U_\mathrm{th})/\sigma_U]}\right).
    \label{eq:newcharnock}
\end{equation}
Here $\hat{\alpha}_0 = 0.0065$ and corresponds to the previously constant $\hat{\alpha}$ in Eq.~(\ref{eq:varcharnock}) while $\hat{\alpha}_{-} = 0.0001$ is the value that $\hat{\alpha}$ asymptotes towards at very high wind speeds. We employ the same parameterization with some minor adjustments. The optimum threshold wind speed $U_\mathrm{th}$ and the transition range $\sigma_U$ over which the transition should take place are not well known. At one extreme \citet{bi2015} found that the drag coefficient starts to level off between $18$ and $27\,\mathrm{m\,s^{-1}}$, while \citet{zweers10} found the drag coefficient to peak around $27\,\mathrm{m\,s^{-1}}$. Others have reported even higher wind speeds for the levelling off \citep{powell03,makin05,holthuijsen12hurricanewaves}. After having tested a range of thresholds between $28$ and $33\,\mathrm{m\,s^{-1}}$ (not shown), we chose $U_\mathrm{th} = 30\,\mathrm{m\,s^{-1}}$, whereas \citet{li2021improving} chose $U_\mathrm{th} = 28\,\mathrm{m\,s^{-1}}$. We set the transition range $\sigma_U = 1 \,\mathrm{m\,s^{-1}}$ in accordance with \citet{wam47r1} and \citet{li2021improving}. The minimum Charnock parameter now takes the functional form shown in Fig.~\ref{fig:charnock}.

\section{The impact of a modified Charnock parameter on the sea state in high-wind situations}
\label{sec:wamruns}
A modified wave model WAM Cycle 4.7 with source terms based on the physics described by \citet{ardhuin10} (see Section~\ref{sec:wam47}) was set up on a 3~km resolution pan-Arctic domain (see Fig.~\ref{fig:map}). In a sub-region that covers the North Sea, the Norwegian Sea and the Barents Sea, high resolution (3~km) surface winds from the NORA3 atmospheric hindcast were used as forcing \citep{haakenstad21nora3}. Lower resolution (approx. 31~km) surface winds from ERA5 \citep{hersbachetal2020}, the latest ECMWF reanalysis, were used in the outer part of the domain, interpolated to a 3~km grid. A linear interpolation was made over a transition zone of 20 grid points inside the boundaries of the smaller domain. The NORA3 high resolution fields were produced by downscaling ERA5 using the non-hydrostatic convection-permitting atmospheric model, HARMONIE-AROME Cy40h1.2 (see \citealt{bengtsson} and \citealt{Seity}). NORA3 includes a surface analysis scheme and is reinitialized from ERA5 every six hours (see \citealt{haakenstad21nora3}), making the two wind fields dynamically consistent.

The WAM model was set up with 30 frequencies logarithmically spanning the range 0.0345 to 0.5476 Hz
and 24 directional bins. WAM was forced with hourly 10~m neutral winds calculated from the NORA3 hindcast and ERA5 as described above, daily ice concentration fields from the ARC-MFC physical reanalysis system of the Copernicus Marine Service (CMEMS), and wave spectra from ERA5 at the boundaries. The model was run in shallow-water mode, i.e., with the full linear dispersion relation, but without depth refraction and bottom-induced breaking. We set $\beta_\mathrm{max} = 1.28$ after having tested a range from $1.25$ to $1.42$ (not shown). 

Two wave model runs covering the period 2011--2012 were carried out. The reference run without modifications to the Charnock parameter is denoted CTRL and ALT is the run with a reduced minimum Charnock parameter above $30\,\mathrm{m\,s}^{-1}$. We have compared the two wave model runs against buoy and platform measurements in locations indicated in Fig.~\ref{fig:map}. 
See Table~\ref{tab:settings} for an overview of the parameter settings used in the CTRL run. In the ALT run the settings are the same, except for the introduction of the reduced minimum Charnock parameter following Eq.~(\ref{eq:newcharnock}).

\begin{figure}[h!]
   \begin{center}
   \includegraphics[width=\linewidth]{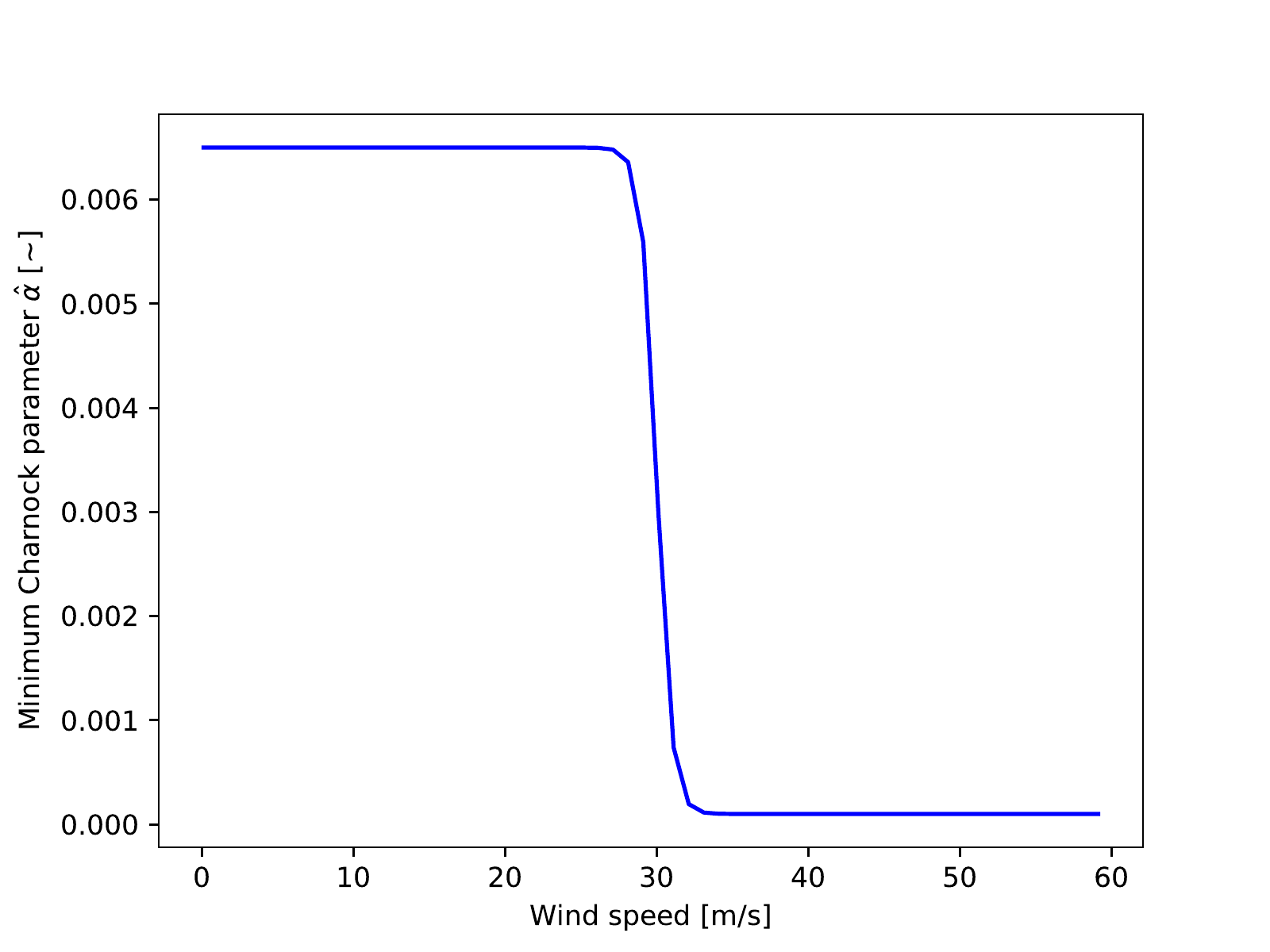}  
   \caption{The tanh formulation, see Eq.~(\ref{eq:newcharnock}), of the high-wind reduction of the minimum Charnock parameter $\hat{\alpha}$ plotted for a threshold wind speed $U_\mathrm{th} = 30.0\,\mathrm{m\,s^{-1}}$ and a transition range $\sigma_U = 1.0\,\mathrm{m\,s^{-1}}$ drag coefficient.}
   \label{fig:charnock}
   \end{center}
\end{figure}

To investigate the impact of the reduction of the Charnock parameter we have compared the drag coefficient and the sea surface roughness in two locations over the period 2011--2012 (Fig.~\ref{fig:cd_alt_ctrl}). The first location is in the region between the Faroe islands and Scotland ($60.51^\circ$N, $006.02^\circ$W, marked as ``+'' just west of K7 in Fig.~\ref{fig:map}). This is an area dominated by synoptic low pressures advancing from the North Atlantic. The fetch in the south-westerly direction is very long. The second location is a grid point near Ekofisk in the central North Sea ($56.48^\circ$N, $003.19^\circ$E, see Fig.~\ref{fig:map}), where the fetch is shorter, except in northerly wind. 
The impact on the drag coefficient, Eq.~(\ref{eq:drag_vs_roughness}), is complex, as it depends on the wave-induced momentum flux $\tau_\mathrm{in}$, see Eq.~(\ref{eq:varcharnock}), and the full history of the wind forcing. Although Eq.~(\ref{eq:newcharnock}) modifies the Charnock parameter significantly only for winds approaching the threshold (here 30~m~s$^{-1}$, see Fig.~\ref{fig:charnock}), the modeled sea state is also affected at weaker winds. This is because, as the wind is decaying, the sea state, and the roughness length, remains high. Thus, the ``memory'' of waves having already seen higher winds is visible below the threshold. This is also referred to as ``old sea'' \citep{hell21swell}. It is thus clear that if stronger winds have been encountered at some point ``up-wave'', the effect will also be seen at winds well below the threshold. In fact, the reduction in the drag coefficient (Fig.~\ref{fig:cd_alt_ctrl}, Panel a) in the ALT run starts to taper off already at 25 m~s$^{-1}$ at the Ekofisk location in the central North Sea (compare the red and the blue curves), whereas in the Faroe-Scotland region the divergence appears at slightly higher wind speeds. This is probably because sea states in the North Sea area are generally younger than in the North Atlantic (also evident from the fact that the drag coefficient is on average lower in the Faroe-Scotland region for high winds). The surface roughness length, $z_0$ (panel b), undergoes a similar flattening and subsequent decrease above wind speeds of 25~m~s$^{-1}$. The drag (and the Charnock parameter, panel c) of the CTRL run is quite high (but more or less in line with what is found by \citealt{li2021improving} for their coupled model) and higher than what is found for coupled systems such as the ECMWF CY41R1 (see, e.g., Fig.~9 by \citealt{pineauguillou18}). The drag in the CTRL run is also higher than what is typically found in measurement campaigns \citep{donelan04,holthuijsen12hurricanewaves,edson13}. 

The wind speed in situ measurements (locations shown in Fig.~\ref{fig:map}) are reduced to 10-m height and compared to the NORA3 hindcast in the upper panel of Fig.~\ref{fig:qqs} together with a quantile-quantile (QQ) comparison.
The two wave model runs are presented in the lower panels. It is evident that the wave growth in the CTRL run (with no modification of the Charnock parameter, see panel b) becomes excessive above 10~m significant wave height. In contrast, the ALT run exhibits a much smaller bias above 10~m significant wave height. It is also clear that the wave field, even for nearly unbiased winds (see the wind comparison for locations K7, Forties, Magnus, and Gullfaks in Fig.~\ref{fig:compFF}, Appendix \ref{app:comp}), is improved in the ALT run (see Fig.~\ref{fig:compHs}). It is particularly interesting to note that there is a small improvement in the Forties location, even though the wind speed rarely exceeds $25\,\mathrm{m\,s}^{-1}$. This must again be related to the memory effect of a sea state which has seen stronger winds, although here presumably in the upwind (``up-wave'') fetch of the Forties location. 

\section{NORA3: A 23-year hindcast archive}
\label{sec:hindcast}
A 23-year wave hindcast covering the period 1998--2020 with settings identical to the ALT run (2011--2012) presented in Section~\ref{sec:wamruns} is presented here (see Appendix~\ref{app:setup} for a list of output parameters and locations of 2D spectra). The inner domain is the North Sea, the Norwegian Sea and the Barents Sea. The resolution, as described above, is 3~km, which coincides with the spatial resolution of the NORA3 atmospheric hindcast \citep{haakenstad21nora3}. The outer part of the domain (forced with ERA5 winds) covers the same domain as the Arctic operational wave forecast model of the Copernicus Marine Service (CMEMS). Boundary spectra are taken from ERA5. 
Our investigation of the model performance is confined to the inner domain.

Fig.~\ref{fig:NORA3stats} presents the 99th percentile significant wave height for the entire period (1998--2020). The mean (Fig.~\ref{fig:map}) and upper percentiles in the open ocean are very similar (see Appendix~\ref{app:nora3_vs_nora10}) but a little weaker than those of the coarser (10~km resolution) NORA10 archive \citep{rei11}. Only in the central North Sea do we see systematic differences in the upper percentiles due to the fact that NORA10 is run in deep-water mode (i.e., no depth-dependent dispersion and refraction). Differences in resolution naturally lead to quite substantial differences in nearshore regions with complex topography where NORA3 is able to resolve more of the coastline and islands and is also able to account for subgrid obstructions (see Appendix~\ref{app:obstructions}). 

Fig.~\ref{fig:statFF} compares the wind speed observations from selected offshore locations against NORA3. \citet{pineauguillou18} found that Jason-2 altimeter measurements of wind speed were biased high compared to buoy measurements in the North Sea and the Norwegian Sea (see their Fig. 4), but tended to yield similar winds to the platform measurements in the North Sea. Our comparison of Sentinel-3A/B, Jason-3 and HaiYang-2B altimeter wind measurements against NORA3 (Fig.~\ref{fig:altimeter}b) suggest on the other hand that the strongest satellite winds are biased a little low compared to the in situ (buoys and platforms) measurements (Fig.~\ref{fig:statFF}).  

\citet{haakenstad21nora3} showed that the 10-m wind at offshore platforms in the North Sea and the Norwegian Sea were in very close agreement with NORA3, and in fact were biased slightly high compared to NORA3 (see their Fig. 9). This was confirmed in the study by \citet{solbrekke21}, who investigated the performance of NORA3 against the FINO-1 wind mast and a number of offshore platforms in the North Sea and the Norwegian Sea. 
The differences between NORA3 and in situ measurements appear to be a little larger in the North Atlantic (see K5 and K7), where we see a slight overestimation of the wind speed above 20~m~s$^{-1}$ for NORA3. We thus cannot rule out that there is a small tendency for a positive upper-percentile bias in the wind speed even if it is not evident in most of the in situ measurements. 

The significant wave height corresponds well to in situ and altimeter measurements all the way up to 14~m (see Fig.~\ref{fig:altimeter}a and Fig.~\ref{fig:statHs}).  
The exceptions are Draugen, Gullfaks and Heidrun, which are known to have radar instruments that are biased low at high wave heights (see the comparison against the NORA10EI hindcast by \citealt{haakenstad20nora10ei}). The significant wave period (zero-crossing period), $T_\mathrm{s}$, shown in Fig.~\ref{fig:statTz}, shows generally good agreement, with correlations in the range 0.82--0.91, but with a small negative bias. This bias is to be expected, since wave buoys have a cutoff frequency of about $0.5~\mathrm{Hz}$, whereas ECWAM and WAM Cycle 4.7 add an $f^{-5}$ spectral tail \citep{wam47r1}. The performance is good throughout the range of wave periods, but certain extreme swell periods are not well captured in the North Atlantic (see location K7).

\section{Discussion and concluding remarks}
\label{sec:discussion}
The wave model WAM Cycle 4.7 has been used to produce a 23-year hindcast forced with wind fields blended from ERA5 and NORA3. The results show good performance in terms of $H_\mathrm{s}$ when the minimum Charnock parameter is reduced above $30\,\mathrm{m\,s}^{-1}$. Its performance is comparable to, or better, than that of the earlier hindcasts NORA10 and NORA10EI \citep{rei11,haakenstad20nora10ei}. 

Accounting for the smoothing of the sea surface by a reduced Charnock parameter is physically motivated, but strongly parameterized. Implementation details will therefore naturally differ depending on the models used, and on the degree of coupling between the models. When applied to the uncoupled NORA3 wave hindcast, the approach yielded results in good agreement with available observations, with only a small upper-percentile bias in the wave heights (Fig.~\ref{fig:altimeter}, top panel, Fig.~\ref{fig:statHs}). For coupled systems, the impact may be even greater, as the near-surface winds respond to a smoother surface by speeding up. This is indeed partly the motivation for the introduction of the modified Charnock parameter in the ECWAM component of ECMWF's IFS \citep{wam47r1}. 

There is virtually no added computational cost associated with the reduction of the Charnock parameter. WAM Cycle 4.7 has also been found to perform well compared against the earlier WAM Cycle 4.5, which used the older WAM physics \citep{bidlot07}. The added cost of introducing the new wave model physics based on \citet{ardhuin10} is also modest (of the order of 14\%, and as source terms are local in physical space, this ratio changes only if the number of frequency and direction bins is changed). Our implementation does not include the cumulative breaking term in the whitecap dissipation presented by \citet{ardhuin10} as it has been found to be very expensive yet having only marginal impact on the sea state \citep{wam47r1}. The implementation, even though we have only tested a standalone wave model here, is efficient enough to be well suited for inclusion in a coupled forecast system, as was shown by \citet{li2021improving} for WAM Cycle 4.7 as well as in the ECMWF ECWAM model \citep{wam47r1}, from which WAM Cycle 4.7 has taken its new model physics.

As (non-hydrostatic) atmospheric models move towards higher horizontal resolution, the upper percentiles of the wind distribution become stronger, but not necessarily too strong. Traditionally, wave models have been tuned to lower winds and their validity must be reconsidered at very high winds. Controlling the wave growth at hurricane-strength winds thus serves the dual purpose of tuning the wave model to stronger winds in a way consistent with observations, and controlling for possible high-wind biases in the atmospheric model. The results appear satisfactory, with small upper-percentile biases against in situ and satellite altimeter wave height measurements (Figs.~\ref{fig:compHs} and \ref{fig:altimeter}a). It is also possible that even when assuming perfect tuning of the wave model, the weak winds typically present in coarser atmospheric models have failed to reveal the deficiencies in the physics since so few high-wind cases have previously appeared. It is quite reasonable to think that too strong wave growth has compensated for weak winds in coarser atmospheric models. The CTRL experiment (Section~\ref{sec:wamruns}, see also Fig.~\ref{fig:compHs}) shows that the impact of the (nearly unbiased) winds in NORA3 is significant as $H_\mathrm{s}$ exceeds 10 m and starts to become really excessive near 15 m. It is also clear that the response of an uncoupled system like NORA3 will be different from that of coupled atmosphere-wave models \citep{wam47r1,li2021improving} where the wind field will adjust to the sea surface roughness and where the Charnock parameter is exchanged, yielding identical roughness for the atmosphere and the wave model \citep{nieuwkoop15}. 

The current high-resolution hindcast also addresses another challenge for high-resolution basin-scale wave models, namely to realistically represent coastal features and open-ocean conditions at the same time. These two regimes are very different \citep{cavaleri18} with strong gradients in winds \citep{christakos21wavegrowth} adding to the complexity in nearshore regions. In our implementation in WAM we try to reconcile these competing needs by pragmatically employing a simple formulation for the reduction of the minimum Charnock parameter, Eq.~(\ref{eq:newcharnock}), and by introducing subgrid obstructions for complex coastlines (see Appendix~\ref{app:obstructions}). It is however clear that more work is needed to further explore the nonlinear processes of short gravity-capillary waves on the wave growth in strong winds \citep{janssen21}, and the simple reduction of the minimum Charnock parameter is clearly a simplification of the response of the sea surface to hurricane strength winds. 

We have shown that including the saturation of the drag can be used to achieve accurate results with low cost in an uncoupled model. It is however clear that further studies into the air-sea momentum balance will require the use of two-way (atmosphere-wave) or even three-way (atmosphere-wave-ocean) coupled models.

\newpage
\begin{table}[]
 \begin{center}
    \begin{tabular}{|l|r|}
       \hline 
       Parameter  & Value \\
       \hline 
       Minimum Charnock parameter $\hat{\alpha}$ & $0.0065$ \\
       Normalized wind input parameter $\beta_\mathrm{max}$ & $1.28$ \\
       Wave age tuning parameter $z_\alpha$ & $0.008$ \\
       Sheltering coefficient $\tau_\mathrm{shelter}$ & $0.25$ \\
       Maximum spectral steepness parameter $\alpha_\mathrm{max}$ & $0.031$  \\
       \hline 
    \end{tabular}
    \caption{Summary of WAM settings used in the CTRL run. The ALT run has identical settings plus a modification of the minimum Charnock parameter.}
    \label{tab:settings}
 \end{center}
\end{table}


\begin{figure}
   \centering
   \includegraphics[width=\linewidth]{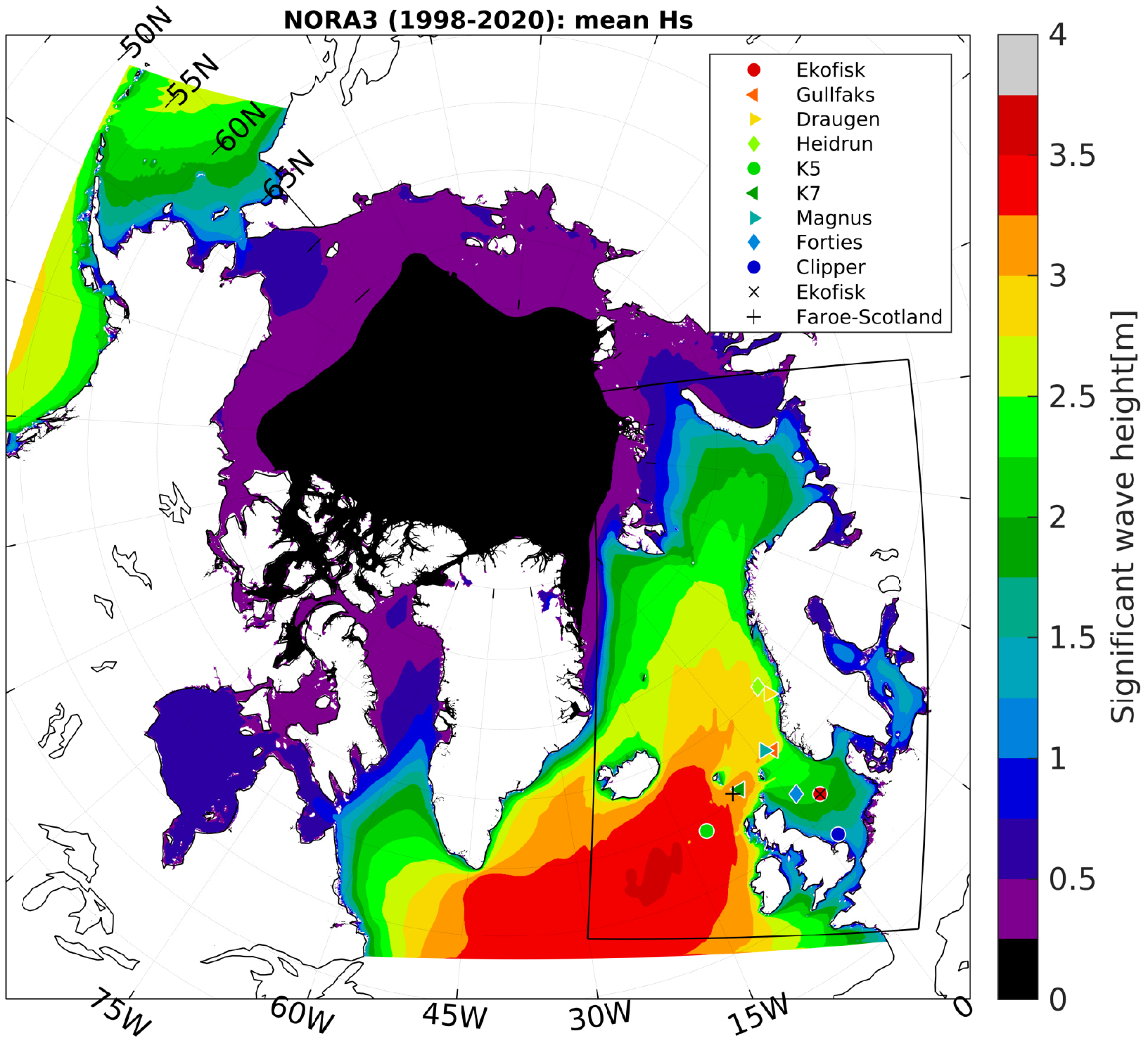}
   \caption{Outline of the WAM model domain with the NORA3 subdomain outlined (black box). The average significant wave height over the period 1998--2020 is shown together with the in situ wave measurement locations. The two locations used for the computation of the drag coefficient and roughness length in Fig.~\ref{fig:cd_alt_ctrl} are indicated with ``x'' and ``+'' for Ekofisk and Faroe-Scotland, respectively.}
   \label{fig:map}
\end{figure}

\begin{figure}
   \centering
   \includegraphics[trim={3.5cm 3.0cm 2.5cm 2.3cm},clip,width=\textwidth]{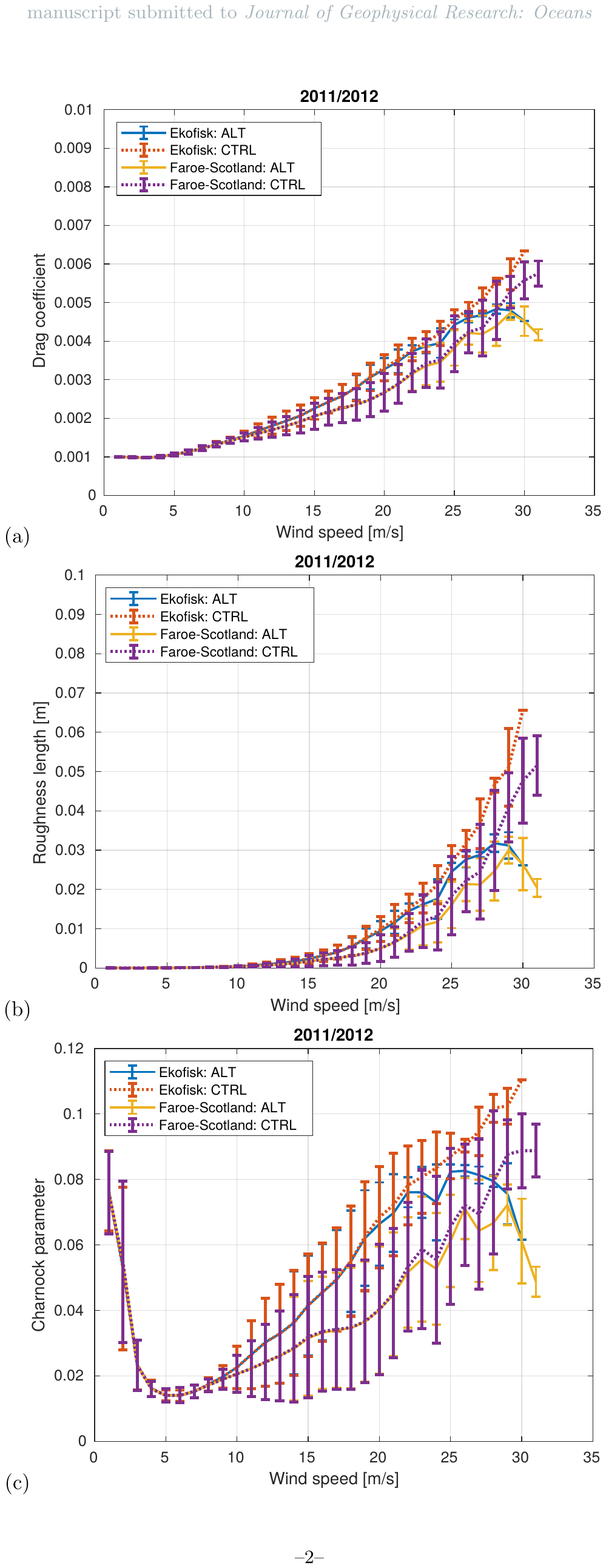} 
   \caption{Drag coefficient $C_\mathrm{D}$ (panel a), roughness length $z_0$ [m] (panel b) and Charnock parameter (panel c) binned at 1 m~s$^{-1}$ resolution and averaged over 2011--2012 for the ALT and CTRL runs. Two locations, the Faroe-Scotland channel and Ekofisk in the central North Sea are shown (locations indicated in Fig.~\ref{fig:map}). Vertical bars represent the standard deviation in each bin.}
   \label{fig:cd_alt_ctrl}
\end{figure}

\begin{figure}
   \centering
   \includegraphics[trim={3.5cm 5.3cm 5.0cm 3.5cm},clip,width=0.8\textwidth]{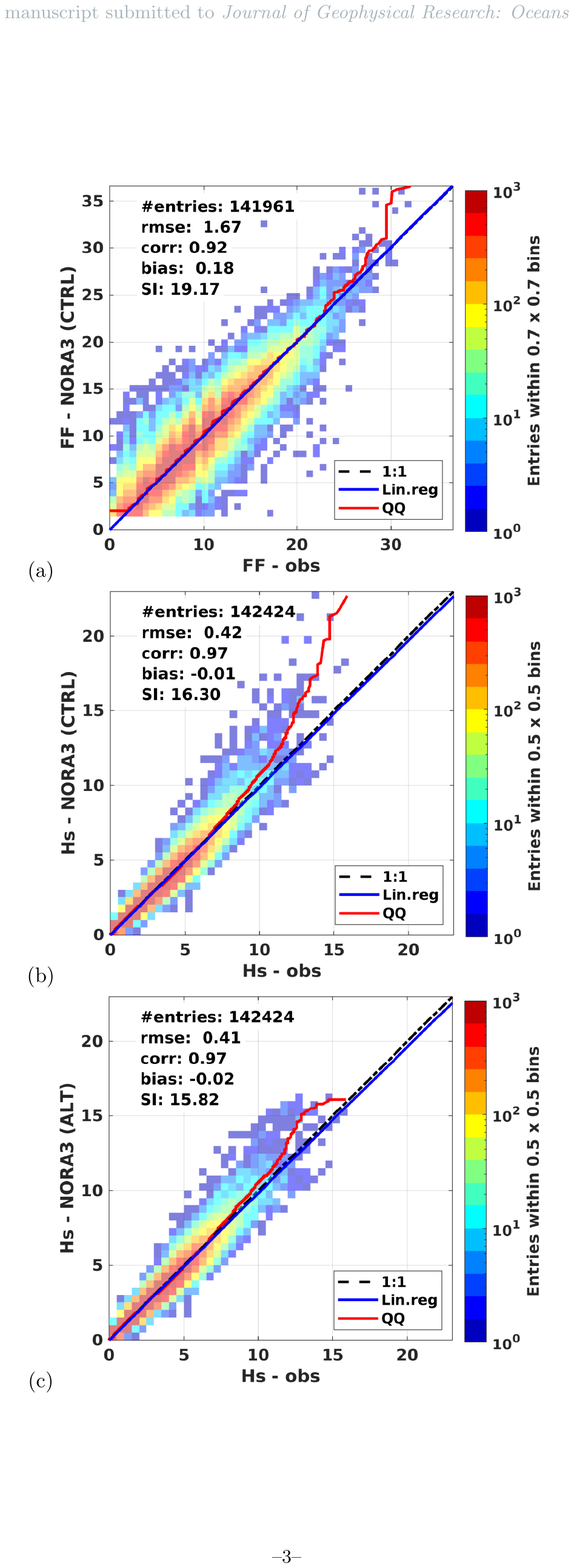} 
   \caption{Aggregate scatter density histograms and quantiles (red line) of all in situ observations in the nine observation locations in Fig.~\ref{fig:map} ($x$-axis) vs. NORA3 ($y$-axis) are shown. Panel a: Comparison of $U_{10}$ [m~s$^{-1}$] wind speed (referred to as FF in the meteorological convention). Panel b Significant wave height of CTRL run, $H_\mathrm{s}$~[m] against observations in locations shown in Fig.~\ref{fig:map}, 2011--2012. Panel c: Same as panel b for ALT run. The number of data points (entries), correlation (cor), bias and regression slope (blue line) are provided in the legend. Quantiles are shown in red. There is a marked reduction in the bias for wave heights above 12~m.}
   \label{fig:qqs}
\end{figure}

\begin{figure}
   \centering
   \includegraphics[width=\linewidth]{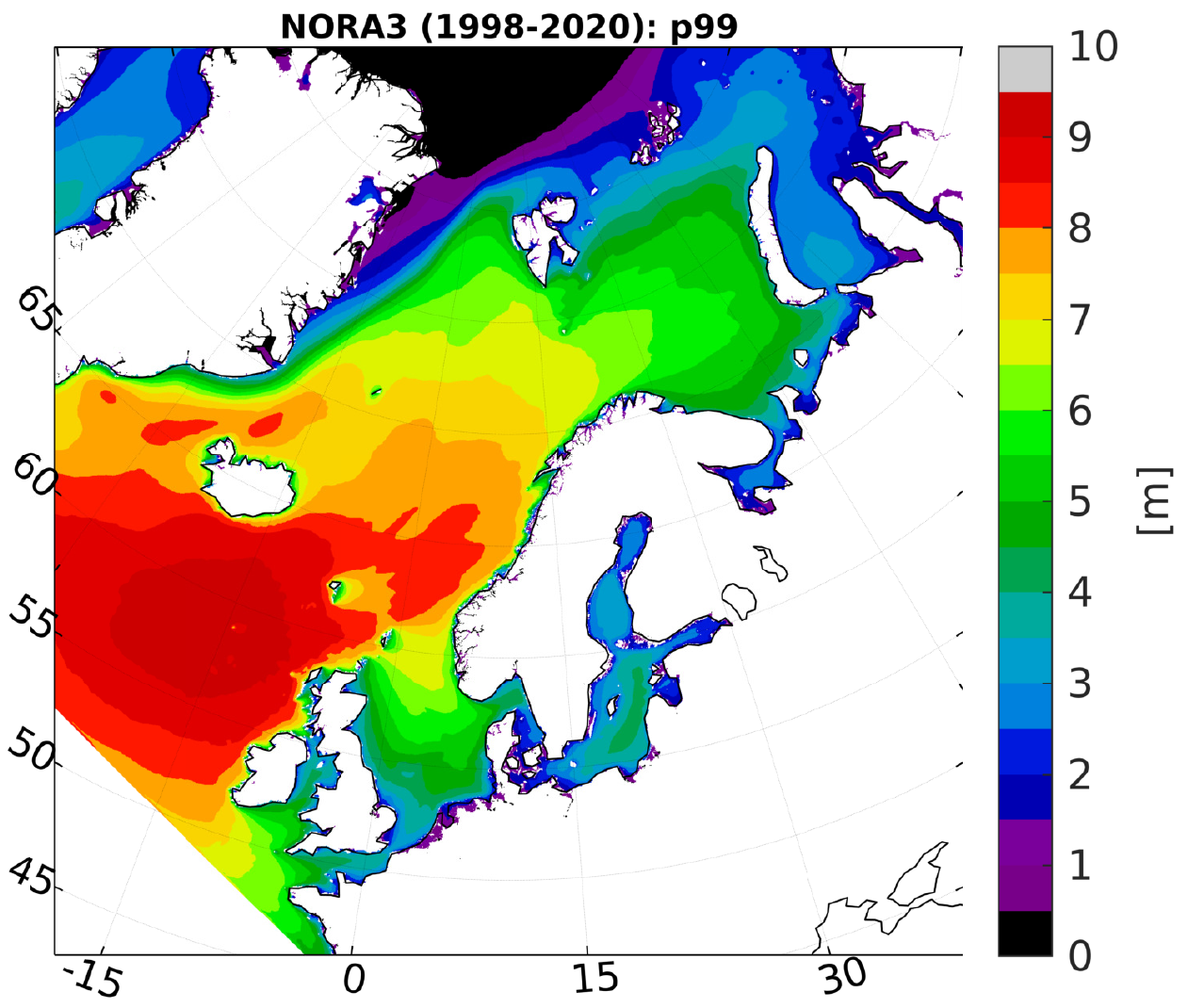}
   \caption{NORA3 WAM 99 percentile statistics (1998--2020) of $H_\mathrm{s}$ [m].}
   \label{fig:NORA3stats}
\end{figure}

\begin{figure}
    \centering
    \includegraphics[trim={3.5cm 8.7cm 1.5cm 7cm},clip,width=\textwidth]{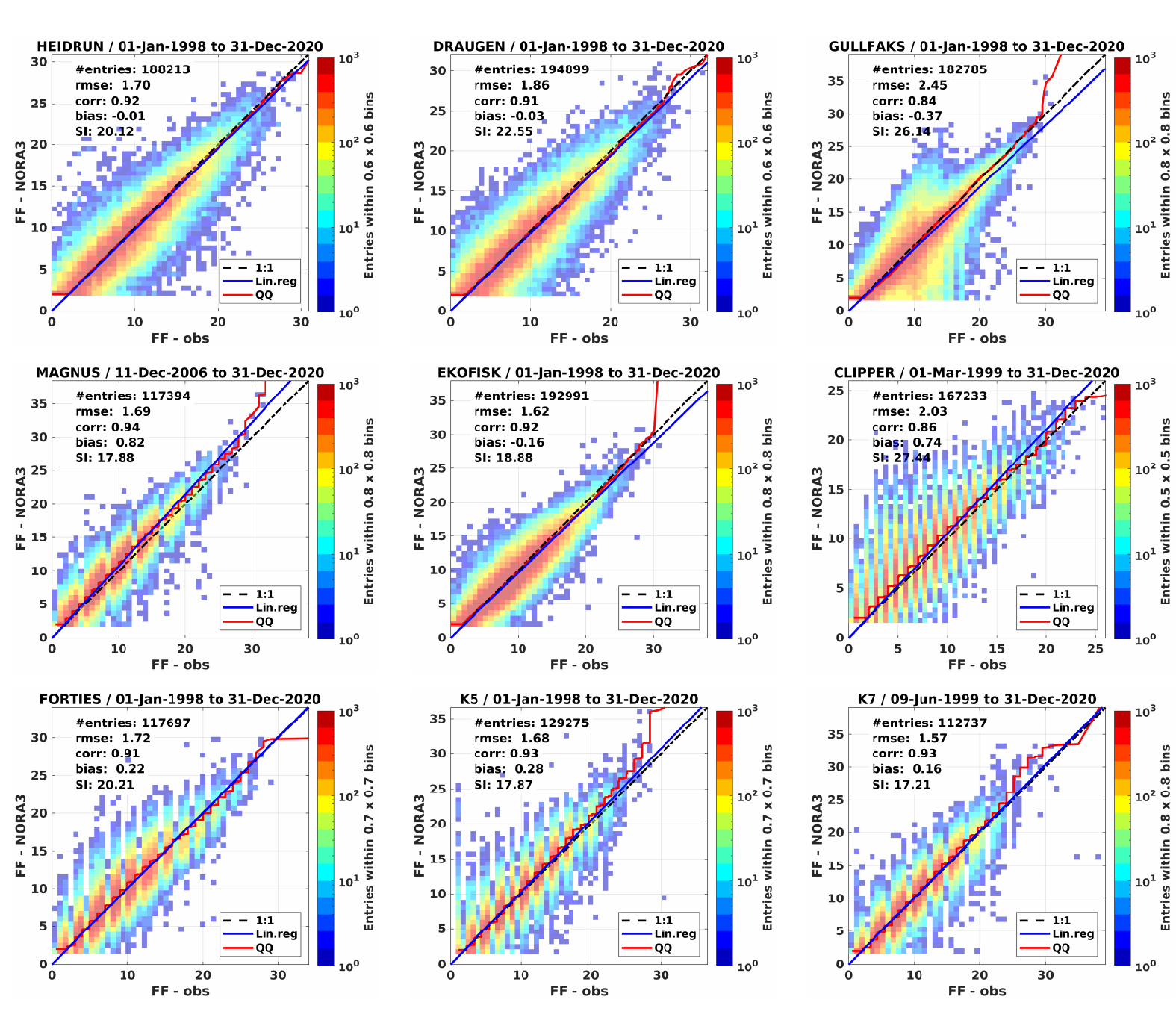} 
\caption{Scatter density histograms and quantiles (red line) of in situ $U_{10} \,[\mathrm{m\,s^{-1}}]$ observations ($x$-axis) vs. NORA3 ($y$-axis). Station and validation period is presented in the title, while the number of corresponding data (entries), correlation (cor), bias and regression slope (blue line) are provided in the legend. Quantiles are shown in red.
\label{fig:statFF}}
\end{figure}

\begin{figure}
   \centering
    \includegraphics[trim={4.5cm 5.2cm 2.5cm 5cm},clip,width=0.95\textwidth]{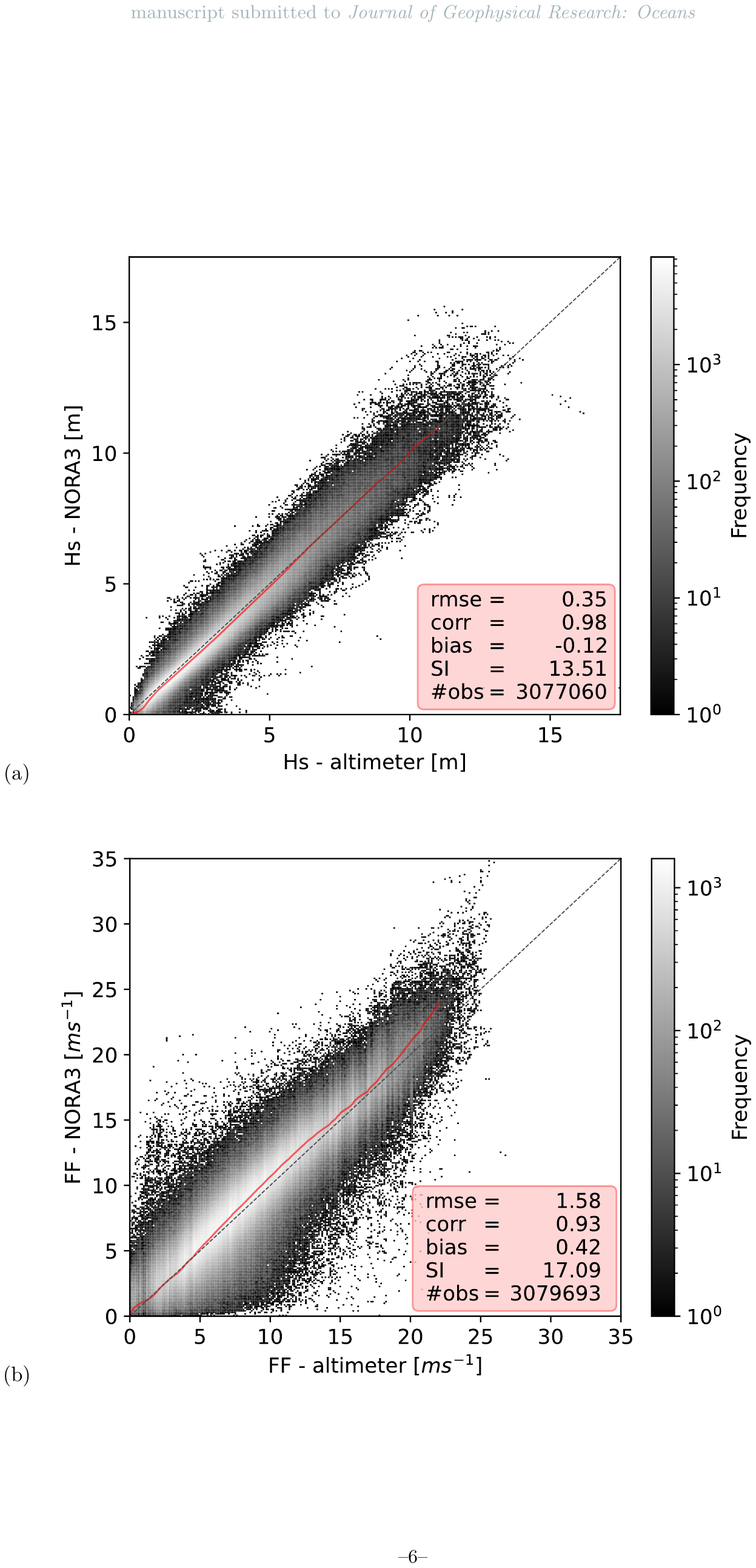} 
   \caption{Panel a: Scatter density histograms and quantiles (red line) of altimeter measurements ($x$ axis) of significant wave height $H_\mathrm{s}\,\mathrm{[m]}$ (Sentinel-3A/B, Jason-3, CFOSAT and HaiYang-2B) vs. NORA3 WAM ($y$ axis). Panel b: Altimeter 10-m wind speed (same satellites as panel a but without CFOSAT) vs. NORA3. All measurements are colocated with model values within the NORA3 WAM subdomain during 2020. The colocation method is described in detail by \cite{bohlinger19novel}. The applied temporal and spatial constraints for the colocation are 30~min and 6~km, respectively. Note that the frequency color scale is logarithmic with outliers plotted as black dots. Quantiles (up to the 99th percentile) are shown in red.}
   \label{fig:altimeter}
\end{figure}

\begin{figure}
    \centering
    \includegraphics[trim={3.5cm 8.7cm 1.5cm 7cm},clip,width=\textwidth]{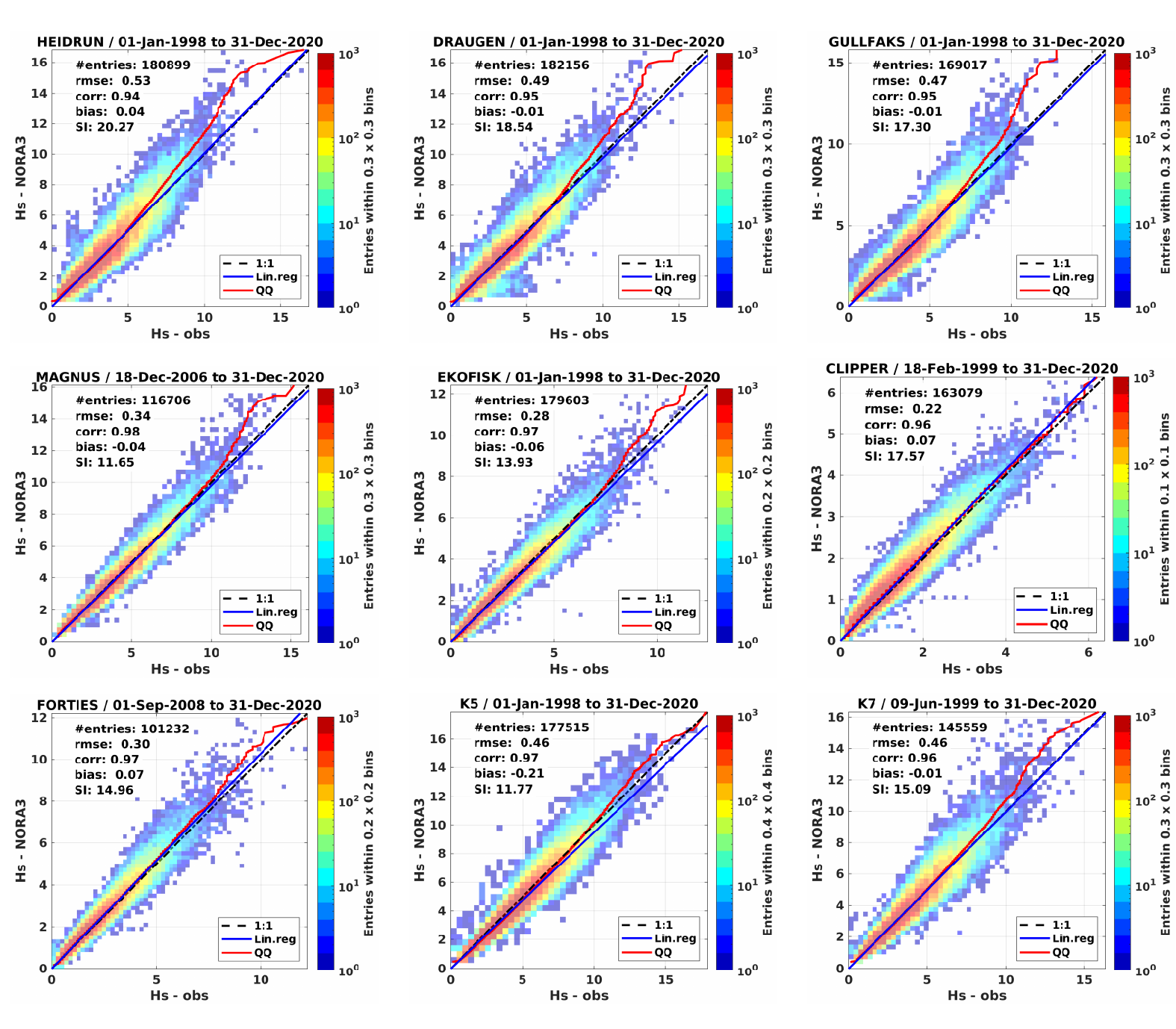} 
\caption{Scatter density histograms and quantiles (red line) of in situ $H_\mathrm{s}\,\mathrm{[m]}$ observations ($x$-axis) vs. NORA3 WAM ($y$-axis). Station and validation period is presented in the title, while the number of corresponding data (entries), correlation (cor), bias and regression slope (blue line) are provided in the legend. Quantiles are shown in red up to the maximum value.
\label{fig:statHs}}
\end{figure}

\begin{figure}
    \centering
    \includegraphics[trim={3.5cm 8.7cm 1.5cm 7cm},clip,width=\textwidth]{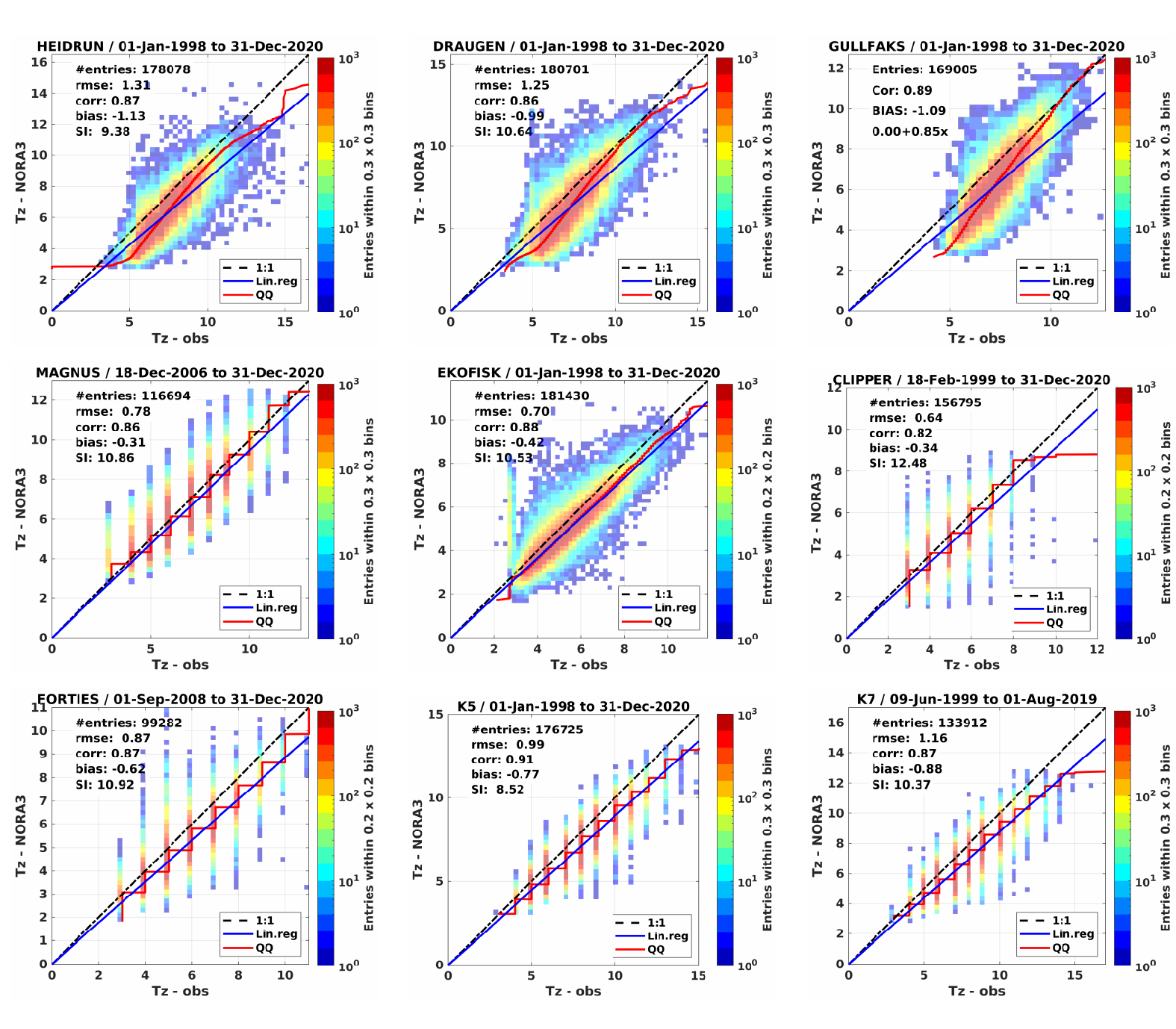} 
\caption{Scatter density histograms and quantiles (red line) of in situ observations of the zero-crossing wave period $T_\mathrm{z}\,\mathrm{[s]}$  ($x$-axis) vs. NORA3 WAM ($y$-axis). Station and validation period is presented in the title, while the number of corresponding data (entries), correlation (cor), bias and regression slope (blue line) are provided in the legend. Quantiles are shown in red up to the maximum value.
\label{fig:statTz}}
\end{figure}

\newpage
\appendix
\section{Comparison of CTRL and ALT hindcast performance at selected offshore and open ocean buoy locations (2011--2012)}
\label{app:comp}
\begin{figure}
    \centering
    \includegraphics[trim={3.5cm 8.7cm 1.5cm 7cm},clip,width=1.2\textwidth]{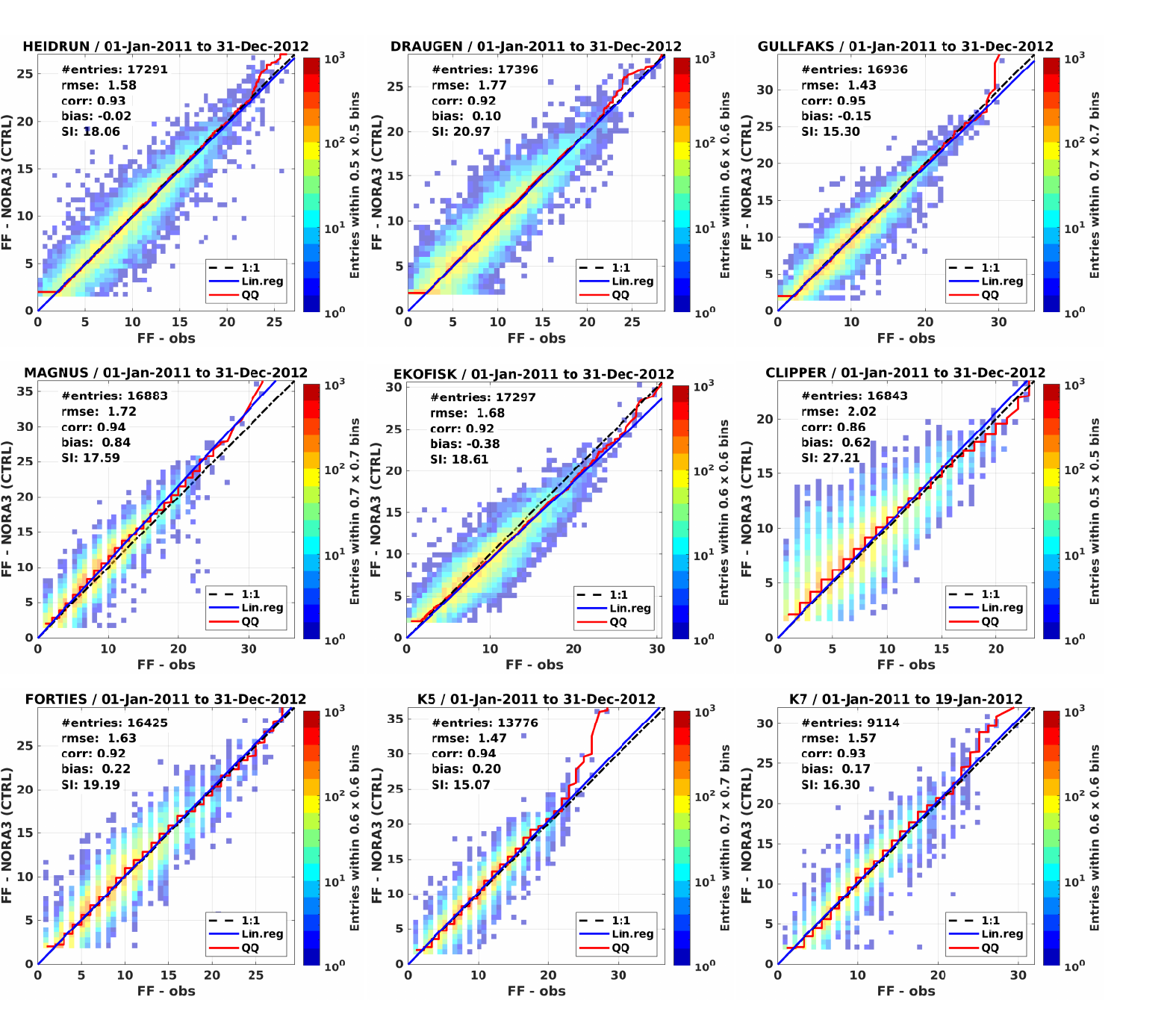} 
\caption{Comparison of $U_{10}\,[\mathrm{m\,s^{-1}}]$, observations ($x$-axis) vs. NORA3 ($y$-axis), 2011--2012. Quantiles are shown in red up to the maximum value.
\label{fig:compFF}}
\end{figure}

\begin{figure}
    \centering
    \includegraphics[trim={3.5cm 8.0cm 2.5cm 6.0cm},clip,width=1.3\textwidth]{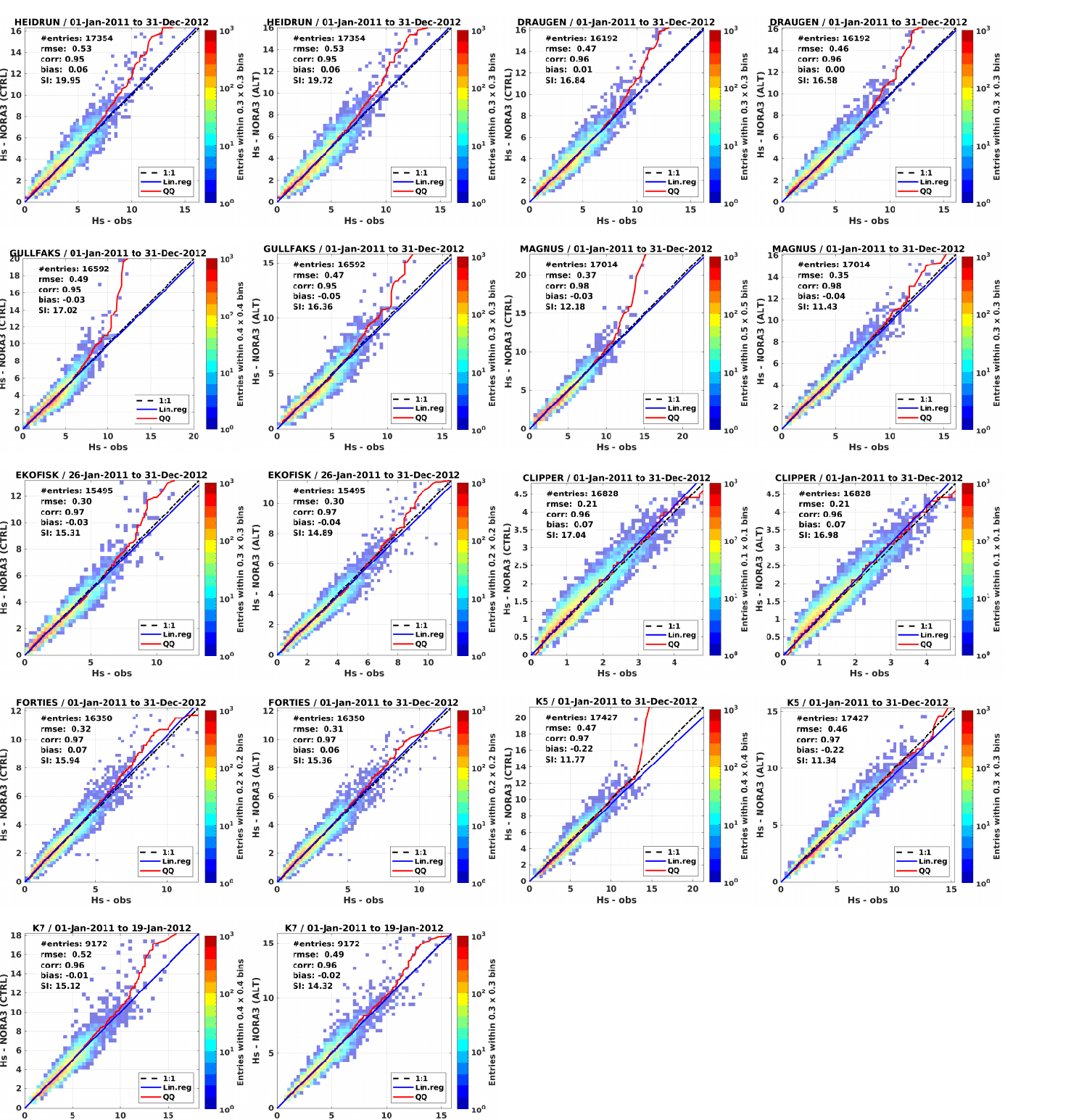} 
\caption{Comparison of $H_\mathrm{s}\,\mathrm{[m]}$, observations ($x$-axis) vs. model ($y$-axis); CTRL (columns 1 and 3) and ALT (columns 2 and 4), 2011--2012. Quantiles are shown in red up to the maximum value.
\label{fig:compHs}}
\end{figure}

The 10~m wind speed measurements for all nine locations shown in Fig.~\ref{fig:map} are compared to the NORA3 wind speed in Fig.~\ref{fig:compFF}. The NORA3 performance is generally found to be good, which is in accordance with what is found by \citet{haakenstad21nora3} and \citet{solbrekke21}, but the uppermost percentiles of K5 and K7 in the North Atlantic are biased a little high. The performance of the CTRL and ALT integrations in terms of significant wave height (2011--2012) is shown in Fig.~\ref{fig:compHs}. Their aggregate performance is shown in Fig.~\ref{fig:qqs}. It is clear that in most locations the upper percentiles in ALT compare better with observations than the reference run (CTRL) with no reduction in drag coefficient. The exception is Clipper which, located close to the east coast of England in the shallow southern North Sea, rarely exceeds a wave height of 5~m.

\section{Subgrid obstructions}
\label{app:obstructions}
The Norwegian coastline is extremely complex (\citealt{adams79}, pp 199--200), and many of its islands, fjords and promontories are not captured even at 3~km resolution. Interpolating from a more detailed topographic data base can also introduce spurious features in the grid, like the closing of narrow inlets or the appearance of unrealistically large islands. We therefore implemented a sub-grid scheme to increase the usefulness of the hindcast for coastal applications. This scheme \citep{tuomi14} uses information from a high-resolution topography to reduce the energy that propagates through partially land-covered grid points. The depth data for the 3~km grid were calculated as the mean of 25 grid points from a 600~m resolution bathymetric grid calculated from the EMODNET topographic database \citep{EMODnetBathymetryConsortium2018EMODnet2018}. The land--sea mask, using a 50\% threshold for land, was then constructed by the method \citet{tuomi14} developed for the Finnish Archipelago Sea. The method is designed to better account for small islands that are spread out over several coarser grid points and therefore may not be detected by simply applying a threshold to each grid point individually. We chose a relatively high threshold because a lower one (e.g., 30\%) can close some of the narrow passages in the fjords. Depths less than 10 m were set as land in the final coarse grid. The method also determined compatible obstruction grids for the coarse sea-points. Finally, adding the sub-grid obstructions to the model run is a standard part in WAM Cycle 4.7, where they are accounted for in the transport equation following \citet{tolman03}.

\begin{figure}
   \includegraphics[trim={3.3cm 9.3cm 3.0cm 8.0cm},clip,width=1.3\textwidth]{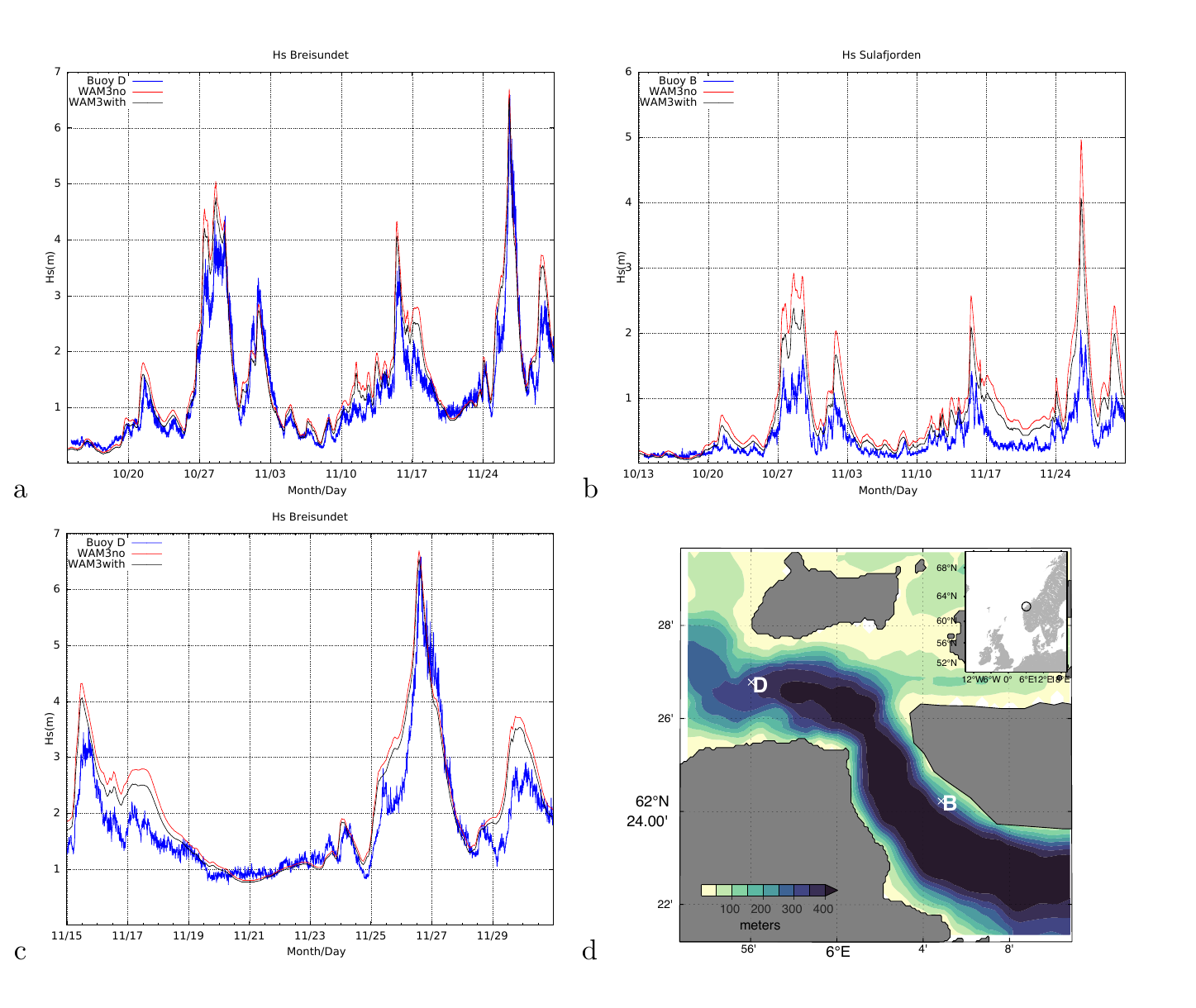} 
   \caption{WAM with and without subgrid obstructions. Panel a: Comparison against buoy D in Breisundet. Panel b: Buoy B further into the fjord shows a larger damping with the subgrid obstructions than the outer buoy. Panel c: Buoy D, close-up of the storm event on 26 November 2016. The NORA3 WAM run with subgrid obstructions is shown in black and the run without obstructions in red, and the observations in blue. Panel d: Map of the Breisundet-Sulafjord area with buoys B and D marked on top of the bathymetry [m]. An inset shows Sulafjord's location on the Norwegian coast.}
   \label{fig:obstructions}
\end{figure}

Nearshore observations are scarce along the coast of Norway with the notable exception of the wave measurement programme operated by the Norwegian Public Roads Administration for the E39 fjord crossing project along the western coast of Norway. The results from two buoys in the Breisundet and Sulafjord area near {\AA}lesund are shown in Fig.~\ref{fig:obstructions} for a period covering October and November 2016. See \citet{christakos20windforcing} and \citet{christakos21wavegrowth} for a detailed account of the observation campaign. Although the differences are mostly modest, it is clear that there is a reduction in the wave height of 10--20~cm in the ALT (hindcast) run with obstructions (black curve, see Panels a, b, and c), compared with the run without obstructions but with all other settings equal to the ALT run (red curve) on the Breisundet buoy (marked ``D'' in Panel d). This has a beneficial impact on the model bias. Panel b shows that buoy ``B'' located further into the fjord exhibits greater damping in the run with obstructions (up to 1 m difference), but the outer buoy (Panels a and c) also experiences some damping due to subgrid obstructions at the mouth of the fjord system. 

\section{Comparison against NORA10 $H_\mathrm{s}$}
\label{app:nora3_vs_nora10}
\begin{figure}
   \includegraphics[trim={3.3cm 3.8cm 3.0cm 2.5cm},clip,width=1.0\textwidth]{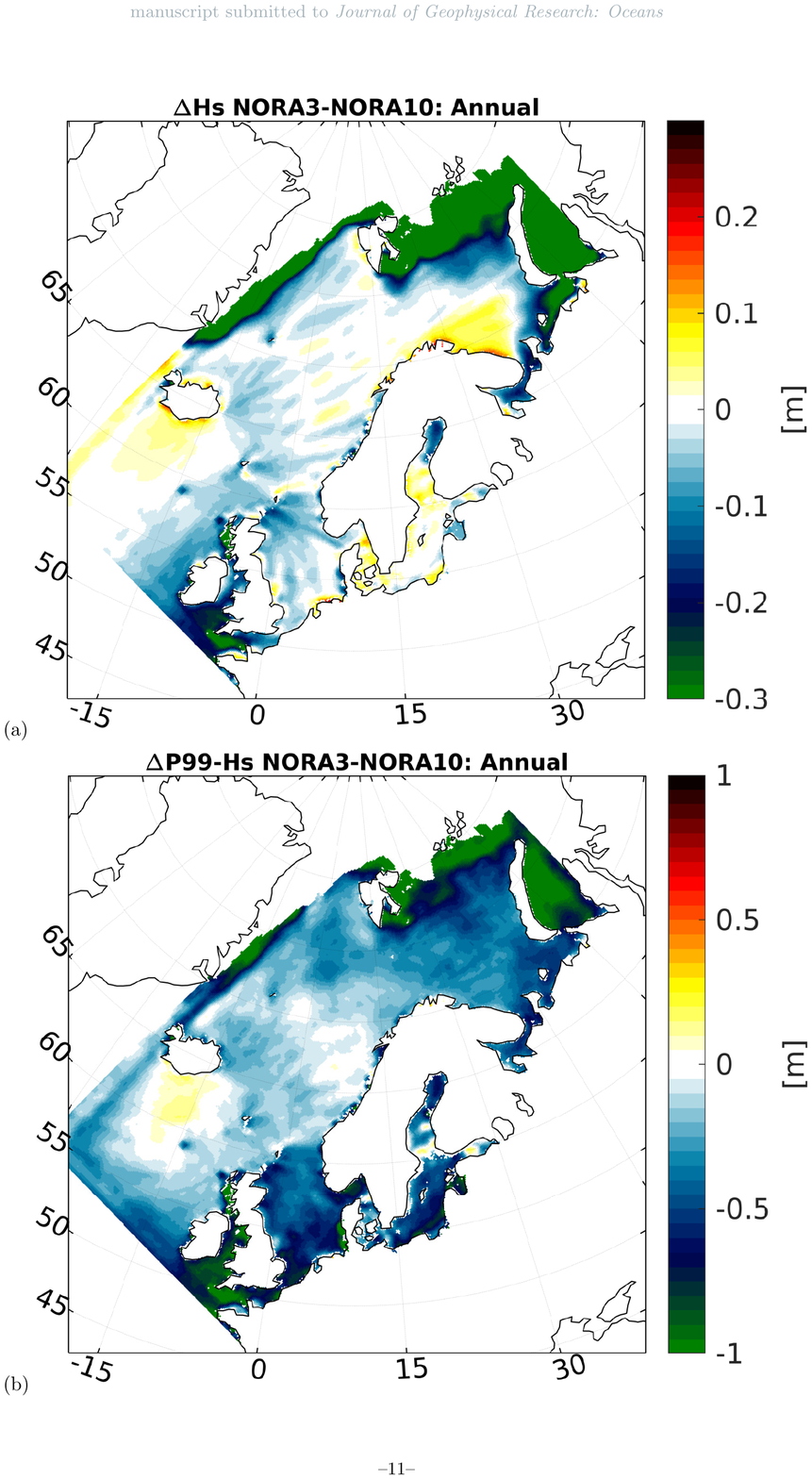} 
   \caption{Difference between NORA3 and NORA10 $H_\mathrm{s}$ for the period 1998--2020. Panel a: Difference in annual average $H_\mathrm{s}$ (NORA3-NORA10). Panel b: Difference in 99th percentile $H_\mathrm{s}$. Note the difference in color scale.}
   \label{fig:nora3_vs_nora10}
\end{figure}

The earlier NORA10 hindcast archive \citep{rei11,aar12} covers approximately the same geographical area as the inner domain of NORA3. Due to its extensive use in extreme value analysis \citep{aar12,bre13b,vanem14} and more generally for mapping the wave climatology of the region \citep{semedo15,bruserud16}, it is of interest to compare the two hindcast archives for the same period (1998--2020). NORA3 has a slightly weaker annual mean $H_\mathrm{s}$ (see Fig.~\ref{fig:nora3_vs_nora10}, panel a), but the differences in the open ocean are of the order of 20~cm. The 99th percentile shows larger differences, as must be expected, but again the differences are small, with most open-ocean regions exhibiting differences of the order of $\pm 0.25$~m. In the North Sea NORA3 is about $0.6$~m lower than NORA10 at the 99th percentile but with very small differences for the annual mean (Panel a). This is partly due to the lack of shallow-water physics in NORA10 which causes the model to overestimate the highest waves in shallow areas. The effect is particularly evident in the shallowest areas in the German Bight. Another effect is the coarser resolution of NORA10 which leads to differences in the shadow between islands. This is particularly evident off the tip of Cornwall and along the west coast of Scotland. Differences in ice extent is the cause of the large differences north of Svalbard and east of Novaya Zemlya.

\section{List of output parameters in the NORA3 WAM wave hindcast}
\label{app:setup}
\begin{figure}
    \centering
   \includegraphics[width=\textwidth]{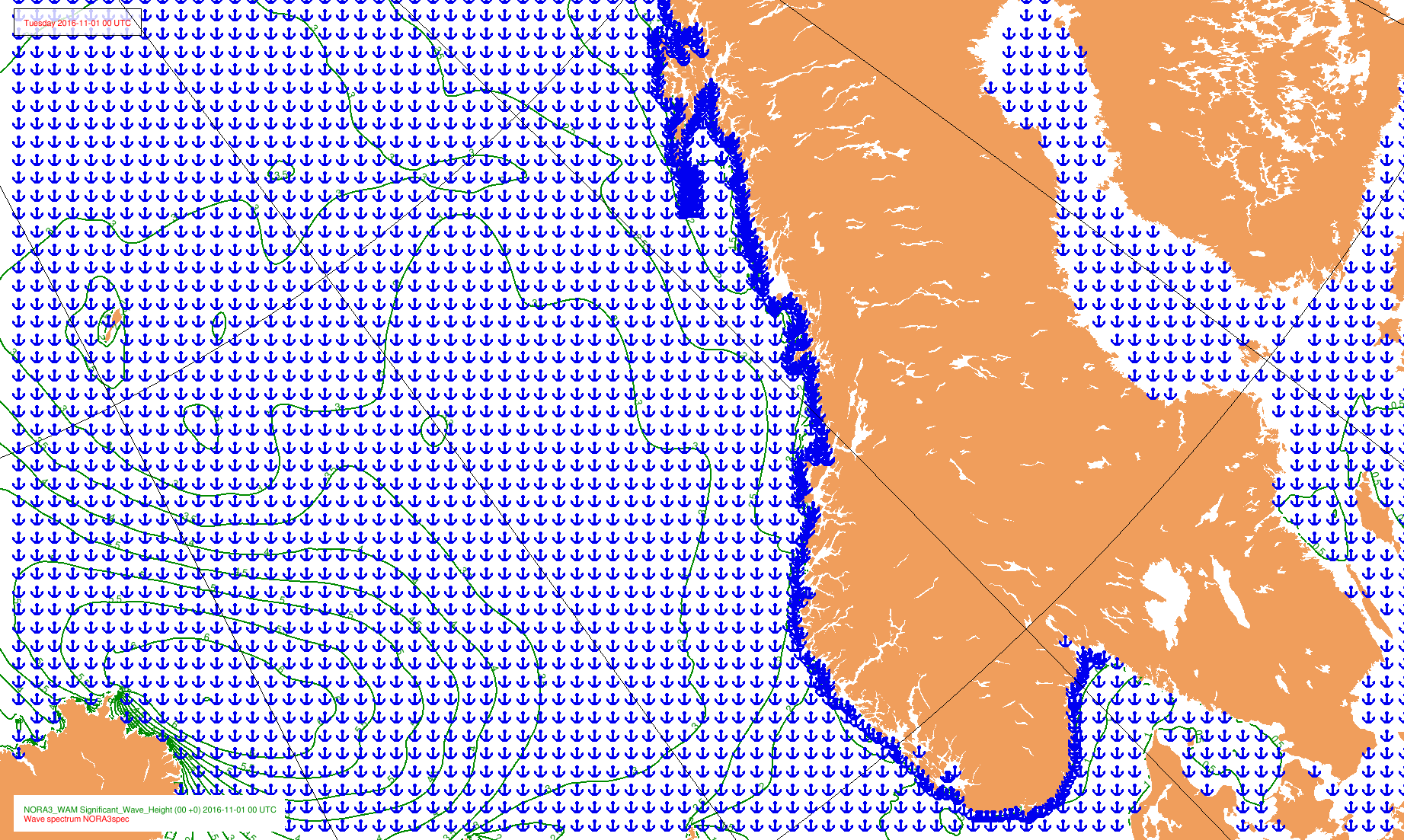}
\caption{Excerpt of the NORA3 WAM model domain showing the typical resolution of the spectral grid. 2D spectra (marked as blue anchors) are archived hourly with a resolution of about $0.25^\circ$ except along the coast of Norway where the resolution is higher.
\label{fig:specloc}}
\end{figure}

Table~\ref{tab:output} shows a truncated output of the contents of the NetCDF files with integrated output parameters from the NORA3 WAM hindcast. All parameters are stored with hourly resolution. 
In addition, the 2D spectra from selected grid points at about $0.25^\circ$ resolution are archived every hour, with higher spatial resolution along the coast of Norway (see Fig.~\ref{fig:specloc}).
\begin{table}
{\tiny
\begin{verbatim}
dimensions:
        time = UNLIMITED  // (24 currently)
        rlat = 1995 
        rlon = 2379
variables:
        rlat: "rotated latitude"; rlat:units = "degrees" 
        rlon: "rotated longitude"; rlon:units = "degrees" 
        ff: "Wind speed"; ff:units = "m/s"
        dd: "Wind direction"; dd:units = "degrees"
        FV: "friction velocity"; FV:units = "m/s" 
        DC: "drag coefficient"; DC:units = "1" 
        hs: "Total significant wave height"; hs:units = "m" 
        tp: "Total peak period"; tp:units = "s" 
        tmp: "Total mean period"; tmp:units = "s" 
        tm1: "Total m1-period"; tm1:units = "s" 
        tm2: "Total m2-period"; tm2:units = "s" 
        thq: "Total mean wave direction"; thq:units = "degrees" 
        hs_sea: "Sea significant wave height"; hs_sea:units = "m" 
        tp_sea: "Sea peak period"; tp_sea:units = "s" 
        tmp_sea: "Sea mean period"; tmp_sea:units = "s" 
        tm1_sea: "Sea m1-period"; tm1_sea:units = "s" 
        thq_sea: "Sea mean wave direction"; thq_sea:units = "degrees" 
        SIC: "sea ice concentration"; SIC:units = "1" 
        hs_swell: "Swell significant wave height"; hs_swell:units = "m" 
        tp_swell: "Swell peak period"; tp_swell:units = "s" 
        thq_swell: "Swell mean wave direction"; thq_swell:units = "degrees" 
        mHs: "expected maximum wave height"; mHs:units = "m" 
        mwp: "expected wave period"; mwp:units = "s" 
        fpI: "interpolated peak frequency"; fpI:units = "s" 
        Pdir: "peak direction"; Pdir:units = "degrees" 
        fshs: "first swell significant wave height"; fshs:units = "m" 
        fstm1: "first swell mean period"; fstm1:units = "s" 
        fsdir: "first swell direction"; fsdir:units = "degrees" 
        sshs: "second swell significant wave height"; sshs:units = "m" 
        sstm1: "second swell mean period"; sstm1:units = "s" 
        ssdir: "second swell direction"; ssdir:units = "degrees" 
        tshs: "third swell significant wave height"; tshs:units = "m" 
        tstm1: "third swell mean period"; tstm1:units = "m/s" 
        tsdir: "third swell direction"; tsdir:units = "degrees" 
        SIT: "sea_ice_thickness"; SIT:units = "m" 
        utrs: "x-comp Stokes drift transport"; utrs:units = "m^2/s" 
        sdx: "x-comp. Stokes drift"; sdx:units = "m/s" 
        sdy: "y-comp. Stokes drift"; sdy:units = "m/s" 
        phioc: "normalized energy flux to ocean"; phioc:units = "W/m^2" 
        phiaw: "normalized energy flux from wind to waves"; phiaw:units = "W/m^2" 
        tauocx: "x-comp. normalized momentum flux into ocean"; tauocx:units = "Pa" 
        tauocy: "y-comp. normalized momentum flux into ocean"; tauocy:units = "Pa" 
        phibot: "energy flux from waves to bottom"; phibot:units = "W/m^2" 
        taubot_x: "x-comp. momentum flux from waves into bottom"; taubot_x:units = "Pa" 
        taubot_y: "y-comp. momentum flux from waves into bottom"; taubot_y:units = "Pa" 
        vtrs: "y-comp Stokes drift transport"; vtrs:units = "m^2/s" 
        projection_ob_tran:grid_north_pole_longitude = 140.0
        projection_ob_tran:grid_north_pole_latitude = 25.0 
        projection_ob_tran:earth_radius = 6371000.0 
        projection_ob_tran:proj4 = "+proj=ob_tran +o_proj=longlat +lon_0=-40 +o_lat_p=25 +R=6.371e+06 +no_defs" 
        latitude:units = "degrees_north" 
        longitude:units = "degrees_east" 
        model_depth: "water depth"; model_depth:units = "m" 
\end{verbatim}}
\caption{List of integrated output parameters from NORA3 WAM.}
\label{tab:output}
\end{table}

\acknowledgments
Thanks go to the editor and the two anonymous reviewers whose constructive comments helped us improve the article. JRB, JS and {\O}B are grateful for support from 
Mercator Ocean and CMEMS through the WaveFlow Service Evolution 2 project, which helped facilitate the implementation of the new physics \citep{ardhuin10} implemented in ECWAM into the open-source WAM Cycle 4.7 and later in WAM Cycle 6.
AC, PB and {\O}B are grateful for support from Mercator Ocean and CMEMS through the ARC-MFC contract. This paper also serves as documentation of the ARC-MFC hindcast archive product \texttt{ARCTIC\_MULTIYEAR\_WAV\_002\_013}. HH and {\O}B gratefully acknowledge the support from ERA4CS through the WINDSURFER project. We are also grateful to Equinor ASA for their continued financial support of the NORA3 hindcast archive as well as the previous NORA10 and NORA10EI archives. We also gratefully acknowledge the Norwegian Public Roads Administration for the long-standing collaboration and their financial support of our research on the Coastal Highway E39 project. We also acknowledge support from the Research Council of Norway through the Stormrisk project (grant no. 300608). The NORA3 wave hindcast is archived and openly available on the THREDDS server of the Norwegian Meteorological Institute: \url{https://thredds.met.no/thredds/projects/windsurfer.html}. The atmospheric hindcast \citep{haakenstad21nora3} is also openly available and archived here: \url{https://thredds.met.no/thredds/catalog/nora3/catalog.html}. The latest version (Cycle 6) of the open-source WAM model (including the subgrid option, but without the reduced high-wind drag option) can be found here: \url{https://github.com/mywave/WAM}.

%
\bibliography{bibliography}


\end{document}